 \documentclass[final,5p,times,twocolumn]{elsarticle}

\usepackage{amssymb}
\usepackage{amsmath}
\usepackage{hyperref}
\usepackage{graphicx}
\usepackage{subcaption}
\usepackage{soul,color,xcolor}

\journal{Physics Letters B}

\begin{document}

\begin{frontmatter}

\title{EGUP corrected thermodynamics of RN-AdS black hole with quintessence matter}

\author{Baoyu Tan\corref{cor1}}
\ead{2022201126@buct.edu.cn}

\address{College of Mathematics and Physics, Beijing University of Chemical Technology,
	15 Beisanhuandonglu Street, Beijing, 100029, China}
\cortext[cor1]{Corresponding author}

\begin{abstract}
Reissner-Nordstr\"{o}m anti de Sitter (RN-AdS) black hole, characterized by electric charge and negative cosmological constant,exhibits a rich thermodynamics structure. In this paper, we consider the influence of quintessence, a hypothetical dark energy component with negative pressure. we have computed the extended generalized uncertainty principle (EGUP) corrections to the thermodynamics of RN-AdS black hole, including Hawking temperature, heat capacity, entropy function and pressure. Furthermore, as a special case of EGUP, we have computed and compared the result obtained from the generalized uncertainty principle (GUP) with those from the extended uncertainty principle (EUP). This work contributes to the understanding of the interplay between fundamental physics and the macroscopic properties of black holes, offering a novel perspective on the thermodynamics of RN-AdS black holes in the context of quintessence and quantum gravity corrections. More importantly, we found that, unlike in the case of the Reissner-Nordstr\"{o}m (RN) black hole, the qualitative behavior for the RN-AdS black hole with quintessence remain largely unchanged, except for minor differences, at the equation of state parameters $\omega_q=-1/3$ and $\omega_q=-2/3$. In addition , unlike RN black holes, the phase transition point of RN-AdS black holes shift to almost zero.
\end{abstract}


\end{frontmatter}

\section{Introduction}
Over the past decades, fundamental advances have been made in the thermodynamics of black holes. Bekenstein showed that for the second law of thermodynamics to hold, black holes should have entropy, and that the entropy of black holes is proportional to the area of the event horizon \cite{PhysRevD.7.2333}. Then, Hawking used quantum field theory of curved space-time to show that black holes have thermal radiation \cite{hawking1975particle}, which can be described by thermodynamic quantities such as temperature, entropy and heat capacity. The black hole entropy formula also contains the fundamental physical constants $G$, $c$, $k$, $h$, which means that black hole thermodynamics embodies the deep and essential relationship between gravitational theory, quantum mechanics and statistical physics.

Black hole thermodynamics is essentially quantum gravitational effects, such as quantum gravity \cite{doi:10.1142/S0217751X95000085}, string theory \cite{KONISHI1990276}, and non-commutative geometry \cite{capozziello2000generalized}, in which clues to a full quantum theory of gravity are hidden. In the last three decades, there have been many exciting new developments in quantum gravity. In particular, t' Hooft and Susskind proposed the holographic principle in the 1990s \cite{hooft2009dimensionalreductionquantumgravity, susskind1995world}. In 1998, Maldacena found a concrete implementation of the holographic principle, the AdS-CFT duality \cite{maldacena1999large}, in which a theory of gravity in an AdS space can be described by a conformal field theory on its boundary. The expended thermodynamics of black holes has also been applied to holographic complexity \cite{PhysRevLett.126.101601}, the weak cosmic supervision conjecture \cite{gwak2017thermodynamics}, and the weak gravity conjecture \cite{RevModPhys.95.035003}. These studies have greatly expanded our understanding of the nature of gravity and space-time.

Quintessence matter (QM) is a scalar field model that describes dark energy \cite{doi:10.1142/S021827180600942X, bahcall1999cosmic, PhysRevD.59.123504}. Kiselev solved the symmetric external solution of the Einstein field equation for a black hole surrounded by QM in Ref. \cite{kiselev2003quintessence}. Some authors have derived the thermodynamics of black holes with QM background \cite{yan2021hawking, huang2021phase, lutfuouglu2021thermodynamics, CHEN2022136994}. At very high energy scales, the Heisenberg uncertainty principle (HUP) is modified, and a minimum measurable distance $(\Delta x)_{min}$ closed to the Planck Length will appear:
\begin{equation}
	\Delta x\Delta p\geq \frac{\hbar}{2}\left(1+\lambda(\Delta p)^2\right),~~~~\lambda>0.
\end{equation}
This is the generalized uncertainty principle (GUP). Here $\lambda$ is a modified parameter such that a finite minimum coordinate uncertainty $(\Delta x)_{min}=\hbar\sqrt\lambda$ occurs. Details of GUP can be found in Ref. \cite{PARK2008698, doi:10.1142/S0218271814300250, FENG201781, doi:10.1142/S0218271817500626, PEDRAM2012638, ali2011minimal, nozari2010minimal, PhysRevD.74.104001, setare2006generalized, maghsoodi2020effect, anacleto2021noncommutative}. Due to the symmetry of the phase space , there may also be minimal uncertainty in the momentum $(\Delta p)_{min}$. Mignemi considered AdS space-time as the topological background of space-time, and extend the HUP by introducting a term proportional to the cosmological constant to show its effects \cite{doi:10.1142/S0217732310033426}. The extended uncertainty principle (EUP)can be written as follow:
\begin{equation}
	\Delta x\Delta p\geq \frac{\hbar}{2}\left(1+\eta(\Delta x)^2\right),~~~~\eta>0.
\end{equation}
The minimum momentum uncertainty that occurs in EUP $(\Delta p)_{min}=\hbar\sqrt\eta$. $\eta$ is proportional to the cosmological constant. Details of EUP can be found in Ref. \cite{CHUNG2019451, hamil2021effect, hamil2021effect1, hamil2021black}. Bolen and Cavagli\'{a} combined EUP and GUP to propose extended generalized uncertainty principle (EGUP), and they considered EGUP as modification of the thermodynamics of Schwarzschild de Sitter black holes \cite{bolen2005anti}.

To this end, we present the structure of this paper here. In section 2, we briefly review EGUP and quintessence. In section 3, we discuss the modification of the thermodynamics of RN-AdS black holes by EGUP and quintessence. Then, we deduce and analyze the cases under the limits of GUP and EUP. We conclude whole paper in section 4. In the following article, we adopt the natural unit system.

\section{A brief review of EGUP and quintessence}
Thermal radiation from black holes is a quantum effect.To describe the quantum behavior of the outgoing particle, the coordinates and momentum of the outgoing particle must satisfy the basic commutation relation of quantum mechanics:
\begin{equation}
	[\hat x_i,\hat p_j]=i\delta_{ij},~~~~i,j=1,2,3.\label{eq:one}
\end{equation}
When we consider the modification of EGUP, the right side of the Eq. (\ref{eq:one}) needs to contain the coordinate operator $\hat x$ and the momentum operator $\hat p$ , which can be expressed by the following formula:
\begin{equation}
	[\hat x,\hat p]=i\left(1+\beta l_p^2\hat p^2+\frac{\alpha}{L^2}\hat x^2\right).\label{eq:two}
\end{equation}
Where $\alpha$, $\beta$ are the modified constants, $l_p$ is the Planck length, and $L$ is the large-scale length. This article only discusses the case where $\alpha>0$ and $\beta>0$. If $\alpha=0$, we find that Eq. (\ref{eq:two}) degenerates into GUP:
\begin{equation}
	[\hat x,\hat p]=i\left(1+\beta l_p^2\hat p^2\right).
\end{equation}
If $\beta=0$, we find that Eq. (\ref{eq:two}) degenerates into EUP:
\begin{equation}
	[\hat x,\hat p]=i\left(1+\frac{\alpha}{L^2}\hat x^2\right).
\end{equation}
In ordinary quantum mechanics, we have the Heisenberg uncertainty principle (HUP):
\begin{equation}
	\Delta x_i\Delta p_j\geq\frac{1}{2}\delta_{ij}.
\end{equation}
Similarly, extended generalized uncertainty principle (EGUP) can be described by the following inequality:
\begin{equation}
	\Delta x\Delta p\geq\frac{1}{2}\left[1+\beta l_p^2(\Delta p)^2+\frac{\alpha}{L^2}(\Delta x)^2\right]\label{eq:three}.
\end{equation}
By solving Eq. (\ref{eq:three}), we find that the uncertainty of momentum $(\Delta p)$ has the following range:
\begin{equation}
	\begin{split}
		\frac{\Delta x}{\beta l_p^2}\left[1-\sqrt{1-\beta l_p^2\left(\frac{1}{(\Delta x)^2}+\frac{\alpha}{L^2}\right)}\right]\leq\Delta p \\ 
		\leq\frac{\Delta x}{\beta l_p^2}\left[1+\sqrt{1-\beta l_p^2\left(\frac{1}{(\Delta x)^2}+\frac{\alpha}{L^2}\right)}\right]\label{eq:four}.
	\end{split}
\end{equation}
Similarly, the coordinate uncertainty $(\Delta x)$ has the following range:
\begin{equation}
	\begin{split}
		\frac{L^2\Delta p}{\alpha}\left[1-\sqrt{1-\frac{\alpha}{L^2}\left(\frac{1}{(\Delta p)^2}+\beta l_p^2\right)}\right]\leq\Delta x \\ 
		\leq\frac{L^2\Delta p}{\alpha}\left[1+\sqrt{1-\frac{\alpha}{L^2}\left(\frac{1}{(\Delta p)^2}+\beta l_p^2\right)}\right]\label{eq:five}.
	\end{split}
\end{equation}
It is not difficult to see, in Eq. (\ref{eq:four}) and Eq. (\ref{eq:five}), the minimum of the momentum uncertainty $(\Delta p)_{min}$ and the minimum of the coordinate uncertainty $(\Delta x)_{min}$ appear naturally.

Einstein gravity coupled to electromagnetic field in the AdS spacetime with quintessence matter can be described by the following action:
\begin{equation}
	S=\int\mathrm{d}^4x~\sqrt{-g}\left[\frac{1}{2}(R-\Lambda)-(\mathcal{L}_{EM}-\mathcal{L}_{QM})\right].
\end{equation}
Where $g$ is the determinant of the metric, $R$ is the curvature scalar, and $\Lambda$ is the cosmological constant. In the AdS spacetime, $l^2=-\frac{3}{\Lambda}$, $l$ is the radius of the AdS spacetime. $\mathcal{L}_{EM}$ is the Lagrangian of the electromagnetic field, which can be given by the following formula:
\begin{equation}
	\mathcal{L}_{EM}=-\frac{1}{4}F^{\mu\nu}F_{\mu\nu}.
\end{equation}
Where $F_{\mu\nu}=\partial_\mu A_\nu-\partial_\nu A_\mu$ is the Faraday tensor of electromagnetic field. Further, the energy-momentum tensor of the electromagnetic field can be obtained:
\begin{align}
	T_{\mu\nu}(EM)&=-\frac{2}{\sqrt{-g}}\frac{\delta(\sqrt{-g}\mathcal{L}_{EM})}{\delta g^{\mu\nu}}\nonumber\\
	&=F_{\lambda\mu}F^{\lambda}{}_{\nu}-\frac{1}{4}F^2g_{\mu\nu}.
\end{align}
Where $F^2=F^{\mu\nu}F_{\mu\nu}$. $\mathcal{L}_{QM}$ is the Lagrangian quantity of quintessence, which can be given by the following formula \cite{doi:10.1142/S0217732321502023, ghosh2018lovelock, PhysRevD.91.123002}:
\begin{equation}
	\mathcal{L}_{QM}=-\frac{1}{2}g^{\mu\nu}\partial_\mu\phi\partial_\nu\phi-V(\phi).
\end{equation}
Further, the energy-momentum tensor of the quintessence can be obtained:
\begin{align}
	T_{\mu\nu}(QM)&=-\frac{2}{\sqrt{-g}}\frac{\delta(\sqrt{-g}\mathcal{L}_{QM})}{\delta g^{\mu\nu}}\nonumber\\
	&=-\partial_\mu\phi\partial_\nu\phi-g_{\mu\nu}V(\phi).
\end{align}
The equation of motion can be obtained from the principle of least action:
\begin{align}
	\frac{1}{\sqrt{-g}}\frac{\delta S(EM+QM)}{\delta g^{\mu\nu}}&=0,\\
	G_{\mu\nu}+\Lambda g_{\mu\nu}&=2T_{\mu\nu}\label{eq:six}.
\end{align}
In Eq. (\ref{eq:six}), we set $\frac{4\pi G}{c^4}=1$, $T_{\mu\nu}=T_{\mu\nu}(EM)+T_{\mu\nu}(QM)$. The line element of RN-AdS spacetime with QM can be solved by Eq. (\ref{eq:six}) as:
\begin{equation}
	\mathrm{d}s^2=-F(r)\mathrm{d}t^2+\frac{1}{F(r)}\mathrm{d}r^2+r^2\mathrm{d}\Omega^2.
\end{equation}
Where $F(r)$ can be expressed as $F(r)=1-\frac{2M}{r}+\frac{Q^2}{r^2}+\frac{r^2}{l^2}-\frac{\chi}{r^{3\omega_q+1}}$, $M$ is the ADM mass of the black hole, $Q$ is the charge carried by the black hole, $l$ is the radius of AdS space-time, satisfying the relation $l^2=-\frac{3}{\Lambda}$, and $\chi$ is the positive normalization factor of QM. For an accelerating universe, the barotropic factor range is $-1\leq\omega_q\leq-\frac{1}{3}$.

In the next section, we will use the previous formula to derive the thermodynamics of RN-AdS black holes modified by EGUP and QM.
\section{EGUP corrected RN-ADS black hole thermodynamics with quintessence}
Let's consider the thermodynamic modification of the RN-AdS black hole with QM by EGUP. First, we get the location of the event horizon from the following formula:
\begin{equation}
	\left.\left(1-\frac{2M}{r}+\frac{Q^2}{r^2}+\frac{r^2}{l^2}-\frac{\chi}{r^{3\omega_q+1}}\right)\right|_{r=r_+}=0\label{eq:seven}.
\end{equation}
According to Eq. (\ref{eq:seven}), the relation between ADM mass $M$ of RN-AdS black hole and the location of its outer event horizon $r_+$ is:
\begin{equation}
	M(r_+)=\frac{r_+}{2}+\frac{Q^2}{2r_+}+\frac{r_+^3}{2l^2}-\frac{\chi}{2r_+^{3\omega_q}}.
\end{equation}

Next, we calculate EGUP's modification of thermodynamic quantities such as temperature, entropy, heat capacity, pressure. So let's start with temperature. Definition of Hawking temperature is \cite{xiang2009heuristic}:
\begin{equation}
	T=\frac{\kappa}{8\pi}\frac{\mathrm{d}A}{\mathrm{d}S}.
\end{equation}
$\kappa$ is the surface gravity, which can be calculated by the metric, the specific expression is:
\begin{align}
	\kappa&=-\frac{1}{2}\lim_{r\to r_+}\sqrt{\frac{g^{11}}{-g_{00}}}g_{00,1}\nonumber\\
	&=\frac{1}{2r_+}\left(1-\frac{Q^2}{r_+^2}+\frac{3r_+^2}{l^2}+\frac{3\omega_q\chi}{r_+^{3\omega_q+1}}\right).
\end{align}
According to the results derived from GUP by Medved and Vagnas \cite{PhysRevD.70.124021}:
\begin{equation}
	\frac{\mathrm{d}A}{\mathrm{d}S}\simeq\frac{(\Delta A)_{min}}{(\Delta S)_{min}}\simeq\frac{\epsilon}{\ln{2}}\Delta x\Delta p.
\end{equation}
Here $(\Delta S)_{min}=\ln{2}$ because the smallest change in entropy is $\ln{2}$. $\epsilon$ is the calibration factor that ensures Hawking's temperature formula returns to the Schwarzschild black hole situation. We assume that the uncertainty of the particle's position is on the order of the diameter of the black hole, $\Delta x\simeq2r_+$. $\Delta p$ is substituted by Eq. (\ref{eq:four}). In order for the result of RN-AdS black hole to return to the Schwarzschild black hole at $Q\to 0$, $\chi\to 0$, $\alpha\to 0$, $\beta\to 0$, $l\to\infty$, $T=\frac{1}{4\pi r_h}$, we get the result that $\epsilon=8\ln{2}$. From the previous discussion, we can see that the Hawking temperature modified by EGUP is:
\begin{align}
	T_{EGUP}=&\frac{2r_+}{\pi\beta l_p^2}\left[1-\sqrt{1-\beta l_p^2\left(\frac{1}{4r_+^2}+\frac{\alpha}{L^2}\right)}\right]\nonumber\\
	&\times\left(1-\frac{Q^2}{r_+^2}+\frac{3r_+^2}{l^2}+\frac{3\omega_q\chi}{r_+^{3\omega_q+1}}\right)\label{eq:eight}.
\end{align}
When $\alpha=0$, we get the GUP-modified Hawking temperature:
\begin{align}
	T_{GUP}=&\frac{2r_+}{\pi\beta l_p^2}\left[1-\sqrt{1-\frac{\beta l_p^2}{4r_+^2}}\right]\nonumber\\
	&\times\left(1-\frac{Q^2}{r_+^2}+\frac{3r_+^2}{l^2}+\frac{3\omega_q\chi}{r_+^{3\omega_q+1}}\right).
\end{align}
When $\beta=0$, we get the EUP-modified Hawking temperature:
\begin{equation}
	T_{EUP}=\frac{r_+}{\pi}\left(\frac{1}{4r_+^2}+\frac{\alpha}{L^2}\right)\left(1-\frac{Q^2}{r_+^2}+\frac{3r_+^2}{l^2}+\frac{3\omega_q\chi}{r_+^{3\omega_q+1}}\right).
\end{equation}
When $\alpha=0$ and $\beta=0$, we get the HUP-modified Hawking temperature:
\begin{equation}
	T_{HUP}=\frac{1}{4\pi r_+}\left(1-\frac{Q^2}{r_+^2}+\frac{3r_+^2}{l^2}+\frac{3\omega_q\chi}{r_+^{3\omega_q+1}}\right).
\end{equation}

\begin{figure}[!h]
	\subfloat[]{\includegraphics[width=7cm]{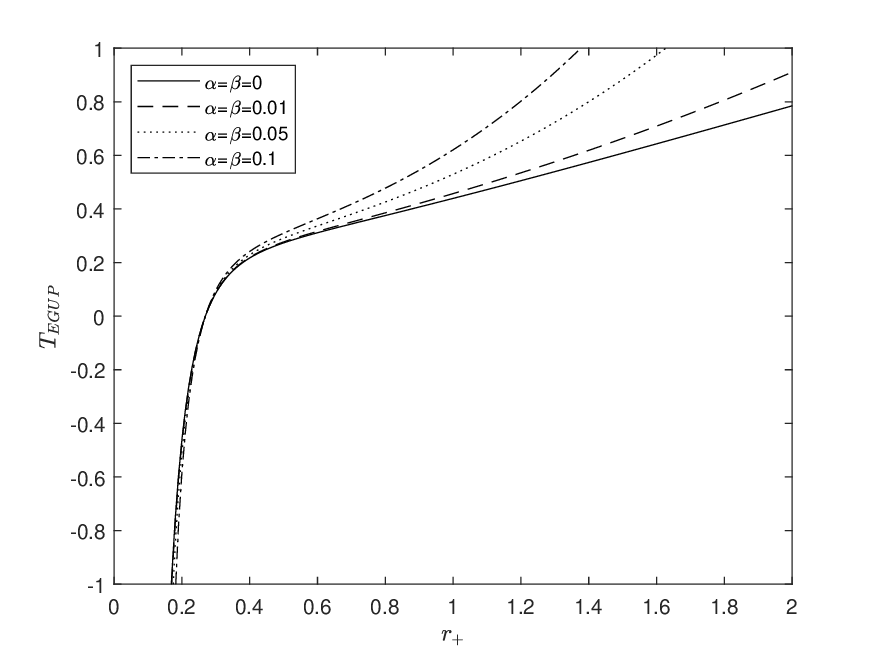}}
	\hfill
	\subfloat[]{\includegraphics[width=7cm]{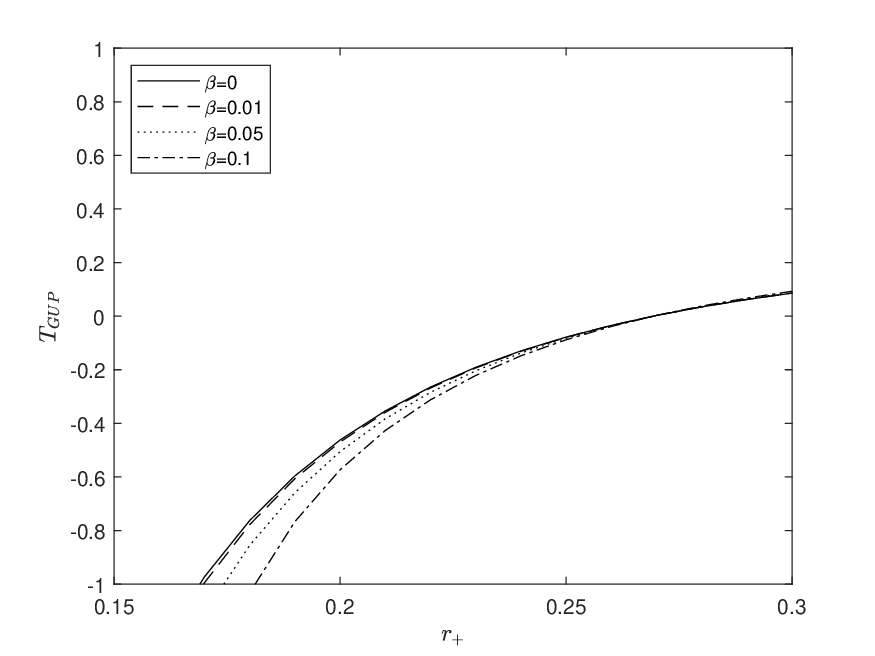}}
	\hfill
	\subfloat[]{\includegraphics[width=7cm]{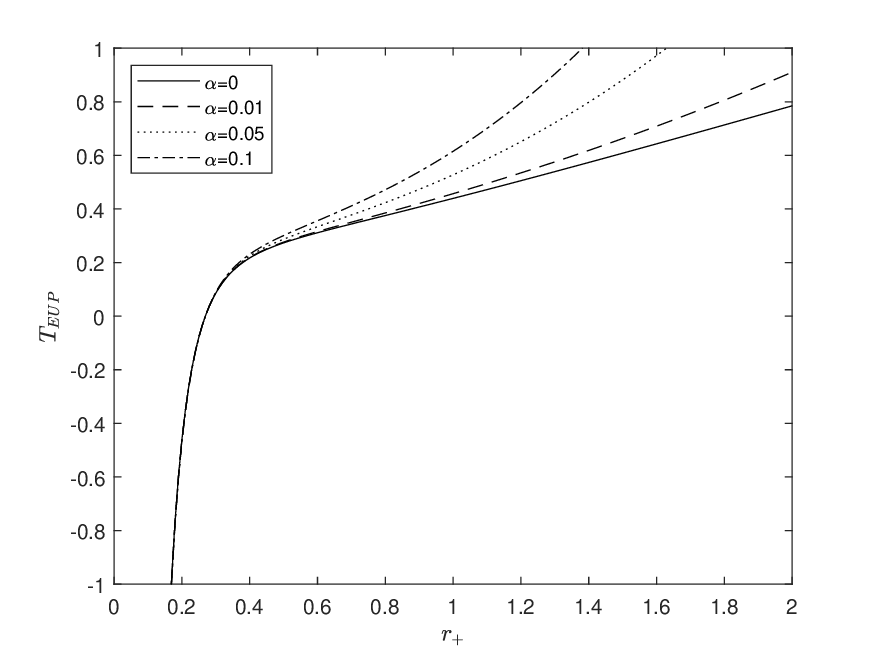}}
	\caption{\label{fig:1}Hawking temperature versus event horizon for $l_p=L=1$, $l^2=\frac{2}{\pi}$, $Q=0.3$, $\chi=0.1$, $\omega_q=-1/3$. (a) EGUP correction of temperature. (b) GUP correction of temperature. (c) EUP correction of temperature.}
\end{figure}

\begin{figure}[!h]
	\subfloat[]{\includegraphics[width=7cm]{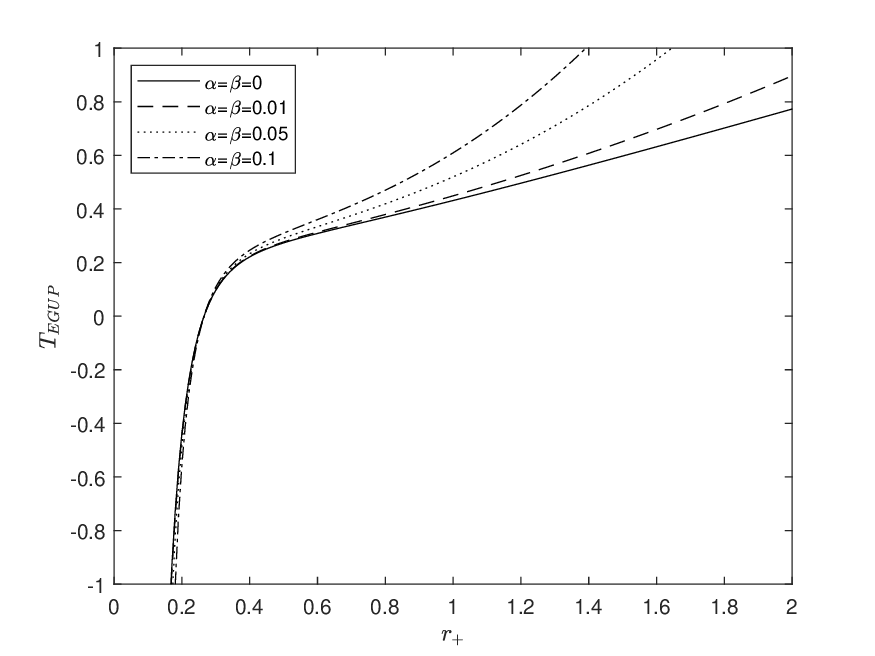}}
	\hfill
	\subfloat[]{\includegraphics[width=7cm]{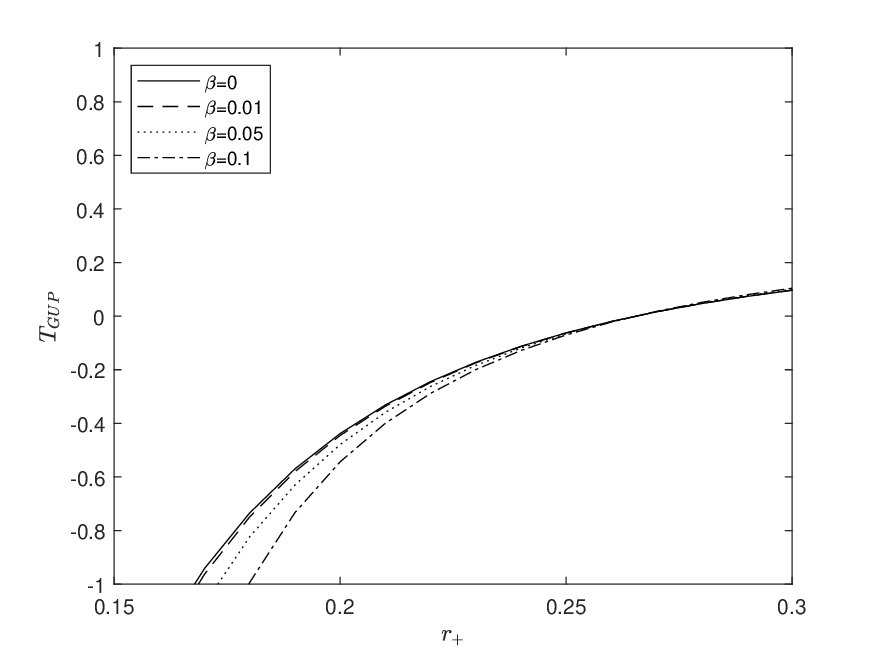}}
	\hfill
	\subfloat[]{\includegraphics[width=7cm]{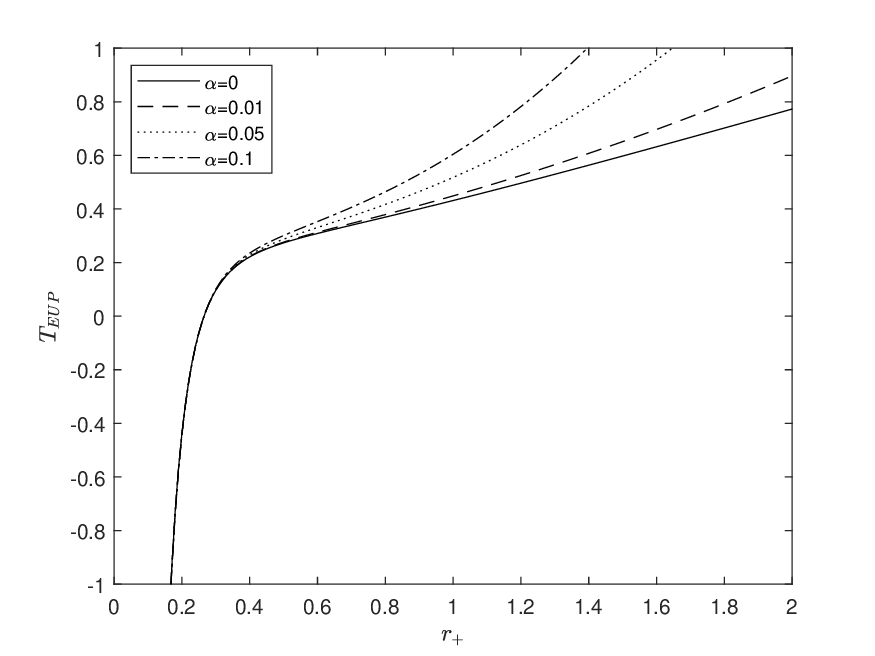}}
	\caption{\label{fig:2}Hawking temperature versus event horizon for $l_p=L=1$, $l^2=\frac{2}{\pi}$, $Q=0.3$, $\chi=0.1$, $\omega_q=-2/3$. (a) EGUP correction of temperature. (b) GUP correction of temperature. (c) EUP correction of temperature.}
\end{figure}

In order to show the modification of EGUP, EUP, and GUP to Hawking temperature in more detail, we show the case of $\omega_q=-1/3$ and the case of $\omega_q=-2/3$ respectively in Fig. \ref{fig:1} and Fig. \ref{fig:2}. In this article, the following values are used for all graphs: $l_p=L=1$, $l^2=\frac{2}{\pi}$, $Q=0.3$, $\chi=0.1$. In addition, in all figures, the GUP modified parameters $\beta$ and the EUP modified parameters $\alpha$ are 0, 0.01, 0.05, 0.1. We found that the GUP correction will lead to a decrease in black hole temperature at small scales, while having almost no effect at large scales. Conversely, the EUP correction has almost no effect at small scales, but leads to an increase in black hole temperature at large scales. The EGUP correction leads to a decrease in black hole temperature at small scales, but an increase at large scales. Additionally, it is worth noting that whether $\omega_q=-1/3$ or $\omega_q=-2/3$ is chosen, there is only a slight correction to the black hole temperature without altering the overall trend.

The Hawking temperature must be both real and positive, which requires Eq. (\ref{eq:eight}) to have some constraints. That is, the last two terms of Eq. (\ref{eq:eight}) must be positive, and the formula under the radical must be positive. For terms containing QM, we only consider the two cases $\omega_q=-1/3$ and $\omega_q=-2/3$. So we have these three inequalities:
\begin{equation}
	0\leq1-\beta l_p^2\left(\frac{1}{4r_+^2}+\frac{\alpha}{L^2}\right)\leq1\label{eq:nine},
\end{equation}
\begin{equation}
	1-\frac{Q^2}{r_+^2}+\frac{3r_+^2}{l^2}-\chi\geq0,~~~~\omega_q=-\frac{1}{3}\label{eq:ten},
\end{equation}
\begin{equation}
	1-\frac{Q^2}{r_+^2}+\frac{3r_+^2}{l^2}-2\chi r_+\geq0,~~~~\omega_q=-\frac{2}{3}\label{eq:eleven}.
\end{equation}
For Eq. (\ref{eq:nine}), under the correction of EUP and HUP, we can only get the same result as before $r_+\geq0$. However, with the modification of EGUP and GUP, we get the minimum that the event horizon of the black hole is non-zero, that is, the black hole is not completely evaporated by thermal radiation. We have the minimum radius with the GUP and EGUP corrections:
\begin{equation}
	r_{+\min}(GUP)=\frac{l_p}{2}\sqrt{\beta}.
\end{equation}
\begin{equation}
	r_{+\min}(EGUP)=\frac{l_pL}{2}\sqrt{\frac{\beta}{L^2-\alpha\beta l_p^2}}.
\end{equation}
Furthermore, we can obtain the minimum mass of a black hole under the corrections of GUP and EGUP.
\begin{equation}
	M_{\min}(GUP)=\frac{l_p}{4}\sqrt{\beta}+\frac{Q^2}{l_p\sqrt{\beta}}+\frac{l_p^3\beta^{3/2}}{16l^2}-\frac{\chi}{2(l_p\sqrt{\beta}/2)^{3\omega_q}}.
\end{equation}
\begin{align}
	M_{\min}(EGUP)=\frac{l_pL}{4}\sqrt{\frac{\beta}{L^2-\alpha\beta l_p^2}}+\frac{Q^2}{l_pL}\sqrt{\frac{L^2-\alpha\beta l_p^2}{\beta}}\nonumber\\
	+\frac{l_p^3L^3}{16l^2}\left(\frac{\beta}{L^2-\alpha\beta l_p^2}\right)^{3/2}-\frac{\chi}{2\left(\frac{l_pL}{2}\sqrt{\frac{\beta}{L^2-\alpha\beta l_p^2}}\right)^{3\omega_q}}.
\end{align}
In the case of Eq. (\ref{eq:ten}), $\omega_q=-1/3$, we find the conditions that must be met for the black hole's outer event horizon $r_+$:
\begin{equation}
	r_+\geq\frac{l}{\sqrt{6}}\sqrt{\chi-1}\sqrt{1+\sqrt{1+\frac{12Q^2}{l^2(\chi-1)^2}}},~~~~0\leq\chi\leq1.
\end{equation}
For Eq. (\ref{eq:eleven}), $\omega_q=-2/3$, the same is obtained:
\begin{align}
	r_{+1}=A-B-C-D,\\
	r_{+2}=A-B+C-D,\\
	r_{+3}=A+B-C+D,\\
	r_{+4}=A+B+C+D.
\end{align}
Where:
\begin{align}
	A=&\frac{l^2\chi}{6},\\
	B=&\frac{1}{2}\sqrt{-\frac{2l^2}{9}+\frac{l^2\chi}{9}+\eta},\\
	C=&\frac{1}{2}\sqrt{-\frac{4l^2}{9}+\frac{2l^4\chi^2}{9}-\eta},\\
	D=&\frac{-\frac{8l^4\chi}{9}+\frac{8l^6\chi^3}{27}}{4\sqrt{-\frac{2l^2}{9}+\frac{l^4\chi^2}{9}+\eta}},\\
	\eta=&\frac{2^{1/3}(l^4-36l^2Q^2)}{9\xi}+\frac{\xi}{9\times2^{1/3}}.
\end{align}
\begin{align}
	\xi=&\sqrt{-4(l^4-36l^2Q^2)^3+(2l^6+216l^4Q^2-108l^6\chi^2Q^2)^2}\nonumber\\
	&\times(2l^6+216l^4Q^2-108l^6Q62\chi^2)^{1/3}.
\end{align}

It is well known that a negative heat capacity means that the black hole cannot exist stably, and only when the heat capacity is positive does it mean that the black hole can exist stably. In other words, only black holes with large enough mass can exist stably, while black holes with small mass are unstable and will evaporate quickly. However, our study seems to show that in RN-AdS spacetime, the phase transition point disappears. And the region with negative heat capacity is very small, low-mass black holes can also exist stably.

In black hole thermodynamics, the heat capacity can be defined in the following way to obtain EGUP's modification of the heat capacity:
\begin{align}
	C_{EGUP}=&\frac{\pi\beta l_p^2}{4(\Theta_1+\Theta_2)}\sqrt{1-\beta l_p^2\left(\frac{1}{4r_+^2}+\frac{\alpha}{L^2}\right)}\times\nonumber\\
	&\left(\frac{1}{2}-\frac{Q^2}{2r_+^2}+\frac{3r_+^2}{2l^2}+\frac{3\omega_q\chi}{2r_+^{3\omega_q+1}}\right).
\end{align}
Where:
\begin{align}
	&\Theta_1=\left(\frac{1}{2}-\frac{Q^2}{2r_+^2}+\frac{3r_+^2}{2l^2}+\frac{9\omega_q^2\chi}{2r_+^{3\omega_q+1}}\right)\times\nonumber\\
	&\left[\sqrt{1-\beta l_p^2\left(\frac{1}{4r_+^2}+\frac{\alpha}{L^2}\right)}+\beta l_p^2\left(\frac{1}{4r_+^2}+\frac{\alpha}{L^2}\right)-1\right],\\
	&\Theta_2=-\frac{\beta l_p^2}{4r_+^2}\left(\frac{1}{2}-\frac{Q^2}{2r_+^2}+\frac{3r_+^2}{2l^2}+\frac{3\omega_q\chi}{2r_+^{3\omega_q+1}}\right).
\end{align}
When $\alpha=0$, the heat capacity correction by GUP is obtained:
\begin{align}
	C_{GUP}=&\frac{\pi\beta l_p^2}{4(\Theta^\prime_1+\Theta^\prime_2)}\sqrt{1-\frac{\beta l_p^2}{4r_+^2}}\times\nonumber\\
	&\left(\frac{1}{2}-\frac{Q^2}{2r_+^2}+\frac{3r_+^2}{2l^2}+\frac{3\omega_q\chi}{2r_+^{3\omega_q+1}}\right).
\end{align}
Where:
\begin{align}
	\Theta^\prime_1=&\left[\sqrt{1-\frac{\beta l_p^2}{4r_+^2}}+\frac{\beta l_p^2}{4r_+^2}-1\right]\times\nonumber\\
	&\left(\frac{1}{2}-\frac{Q^2}{2r_+^2}+\frac{3r_+^2}{2l^2}+\frac{9\omega_q^2\chi}{2r_+^{3\omega_q+1}}\right),\\
	\Theta^\prime_2=&-\frac{\beta l_p^2}{4r_+^2}\left(\frac{1}{2}-\frac{Q^2}{2r_+^2}+\frac{3r_+^2}{2l^2}+\frac{3\omega_q\chi}{2r_+^{3\omega_q+1}}\right).
\end{align}
When $\beta=0$, the heat capacity correction by EUP is obtained:
\begin{equation}
	C_{EUP}=\frac{\pi}{4(\Theta^{\prime\prime}_1+\Theta^{\prime\prime}_2)}\left(\frac{1}{2}-\frac{Q^2}{2r_+^2}+\frac{3r_+^2}{2l^2}+\frac{3\omega_q\chi}{2r_+^{3\omega_q+1}}\right).
\end{equation}
Where:
\begin{equation}
	\Theta^{\prime\prime}_1=\frac{1}{2}\left(\frac{1}{4r_+^2}+\frac{\alpha}{L^2}\right)\left(\frac{1}{2}-\frac{Q^2}{2r_+^2}+\frac{3r_+^2}{2l^2}+\frac{9\omega_q^2\chi}{2r_+^{3\omega_q+1}}\right),
\end{equation}
\begin{equation}
	\Theta^{\prime\prime}_2=-\frac{1}{4r_+^2}\left(\frac{1}{2}-\frac{Q^2}{2r_+^2}+\frac{3r_+^2}{2l^2}+\frac{3\omega_q\chi}{2r_+^{3\omega_q+1}}\right).
\end{equation}
When $\alpha=0$ and $\beta=0$, the heat capacity correction by HUP is obtained:
\begin{equation}
	C_{HUP}=\pi r_+^2\frac{\frac{1}{2}-\frac{Q^2}{2r_+^2}+\frac{3r_+^2}{2l^2}+\frac{3\omega_q\chi}{2r_+^{3\omega_q+1}}}{-\frac{1}{4}+\frac{3Q^2}{4r_+^2}+\frac{3r_+^2}{4l^2}+\frac{3\omega_q(2-3\omega_q)\chi}{4r_+^{3\omega_q+1}}}.
\end{equation}
When $Q\to 0$, $\chi\to 0$, $l\to\infty$, go back to the Schwarzschild black hole situation: $C=-2\pi r_h^2$.

\begin{figure}[!h]
	\subfloat[]{\includegraphics[width=7cm]{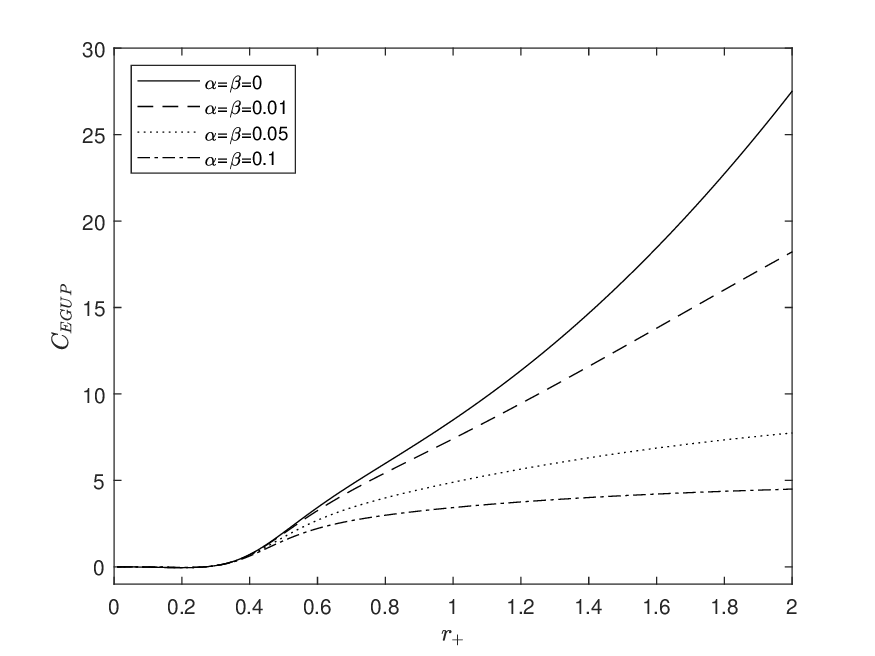}}
	\hfill
	\subfloat[]{\includegraphics[width=7cm]{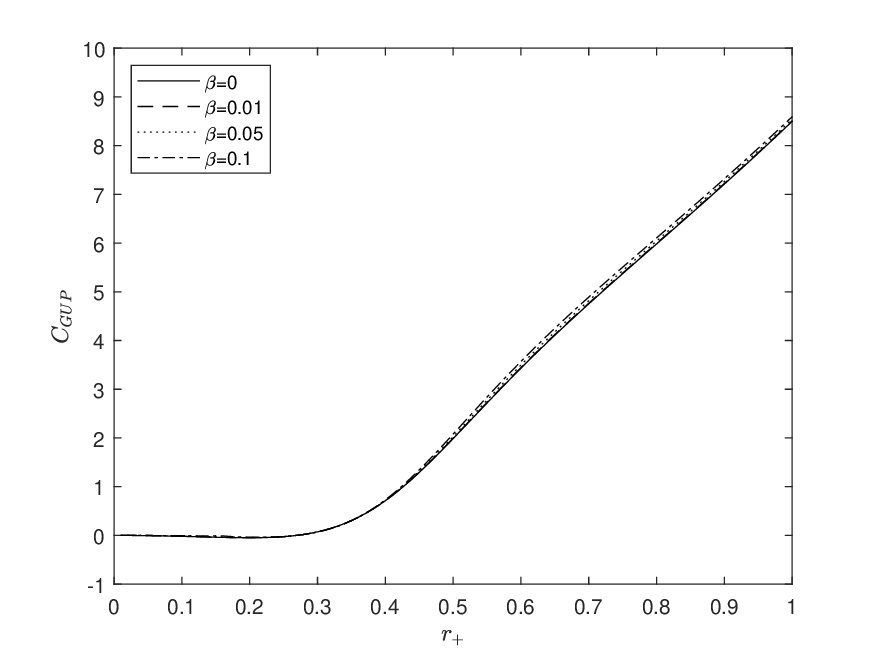}}
	\hfill
	\subfloat[]{\includegraphics[width=7cm]{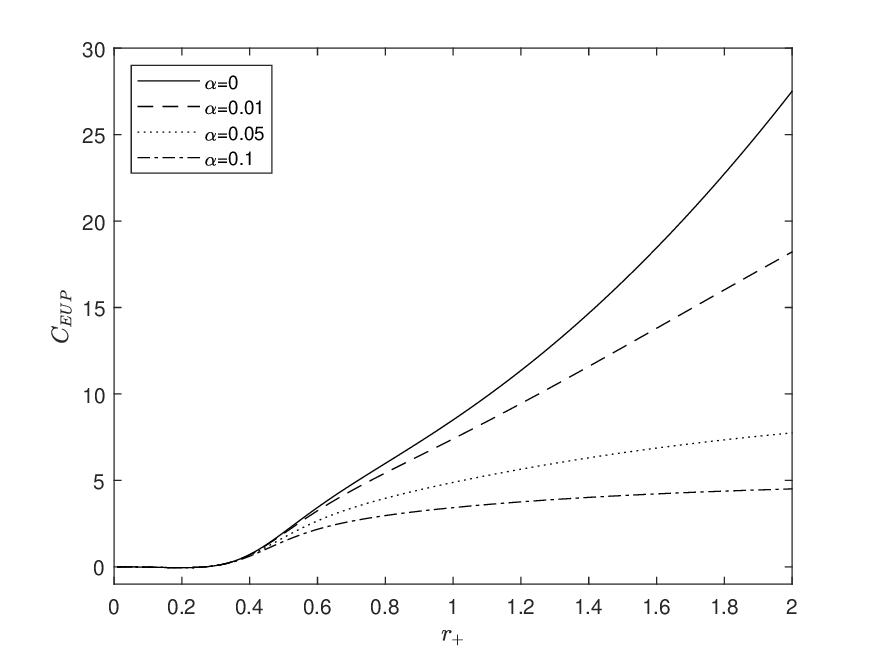}}
	\caption{\label{fig:3} Heat capacity function versus event horizon for $l_p=L=1$, $l^2=\frac{2}{\pi}$, $Q=0.3$, $\chi=0.1$, $\omega_q=-1/3$. (a) EGUP correction of heat capacity. (b) GUP correction of heat capacity. (c) EUP correction of heat capacity.}
\end{figure}

\begin{figure}[!h]
	\subfloat[]{\includegraphics[width=7cm]{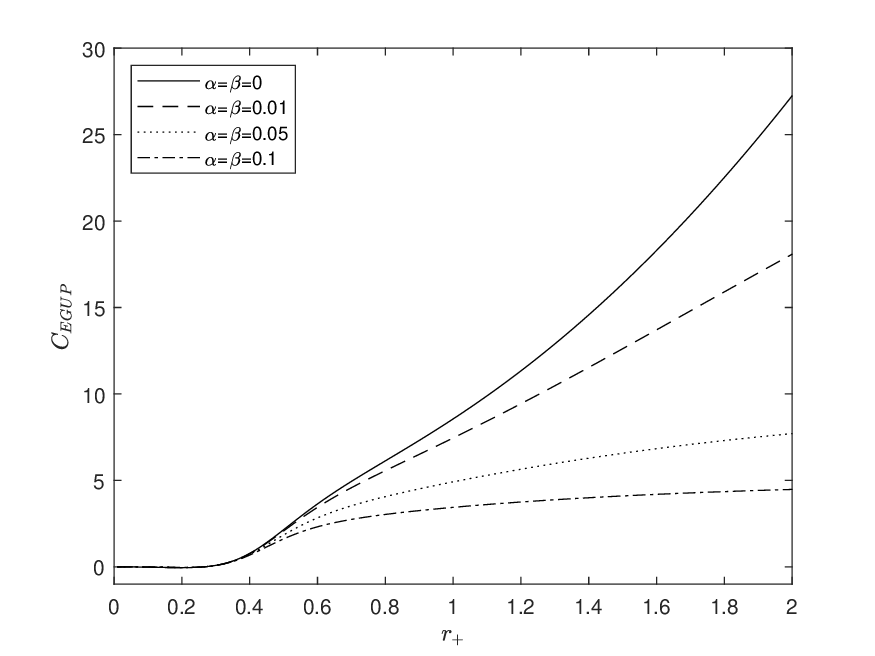}}
	\hfill
	\subfloat[]{\includegraphics[width=7cm]{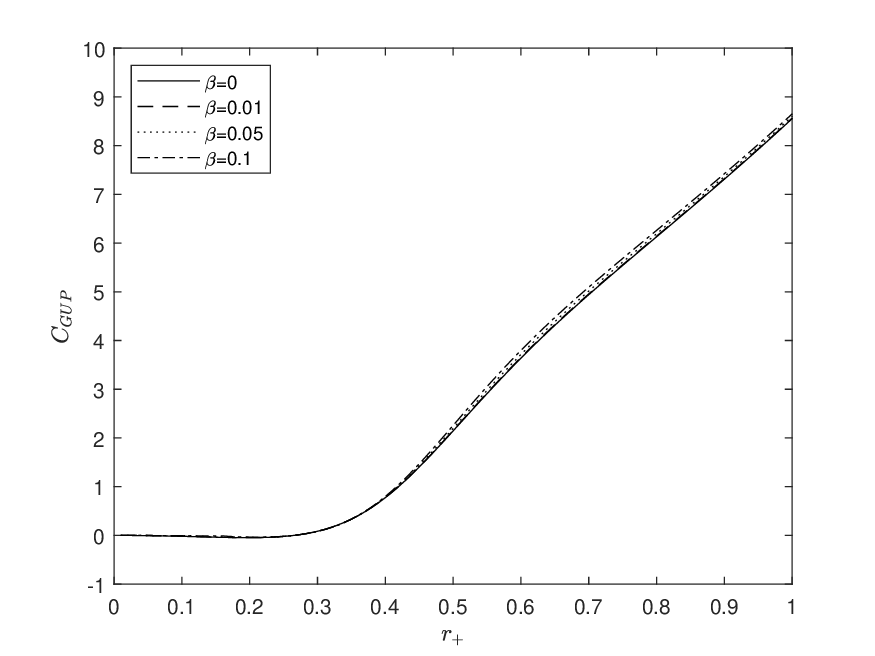}}
	\hfill
	\subfloat[]{\includegraphics[width=7cm]{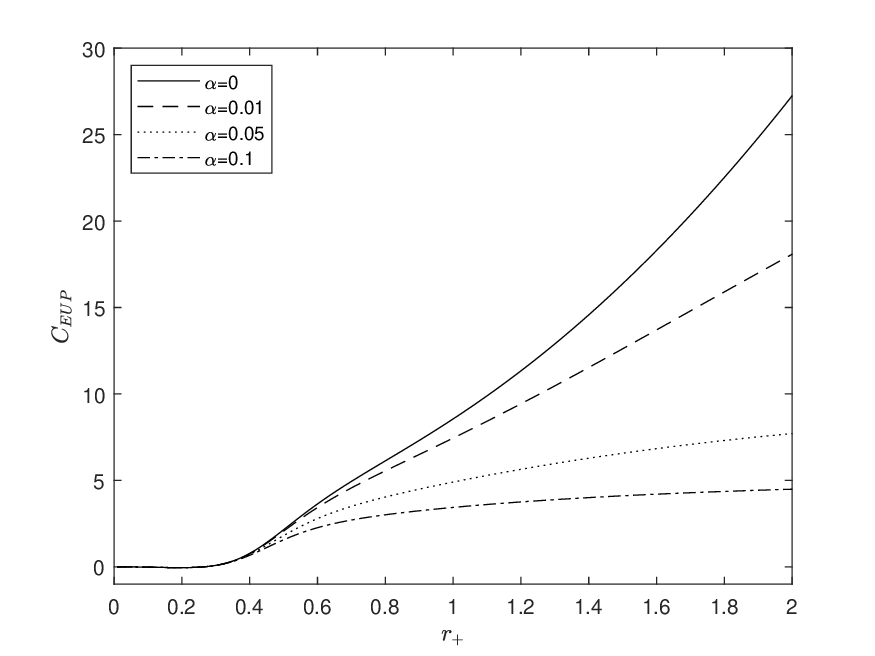}}
	\caption{\label{fig:4} Heat capacity function versus event horizon for $l_p=L=1$, $l^2=\frac{2}{\pi}$, $Q=0.3$, $\chi=0.1$, $\omega_q=-2/3$. (a) EGUP correction of heat capacity. (b) GUP correction of heat capacity. (c) EUP correction of heat capacity.}
\end{figure}

In order to more clearly show the change of heat capacity with the position of the black hole event horizon, we draw the case of $\omega_q=-1/3$ and the case of $\omega_q=-2/3$ in Fig. \ref{fig:3} and Fig. \ref{fig:4} respectively. As can be seen from the figure, GUP correction has a small impact on heat capacity, but EUP correction has a large impact on heat capacity. We found that regardless of whether we take $\omega_q=-1/3$ or $\omega_q=-2/3$, the GUP correction slightly increases the heat capacity of the black hole, while the EUP and EGUP corrections significantly decrease the heat capacity of the black hole at large scales.

\begin{figure}[!h]
	\subfloat[]{\includegraphics[width=7cm]{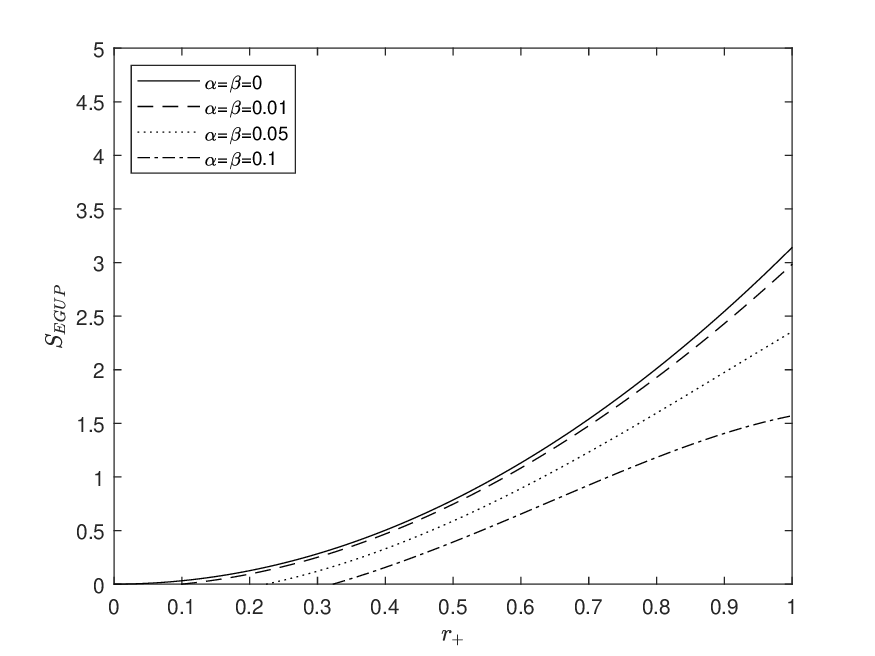}}
	\hfill
	\subfloat[]{\includegraphics[width=7cm]{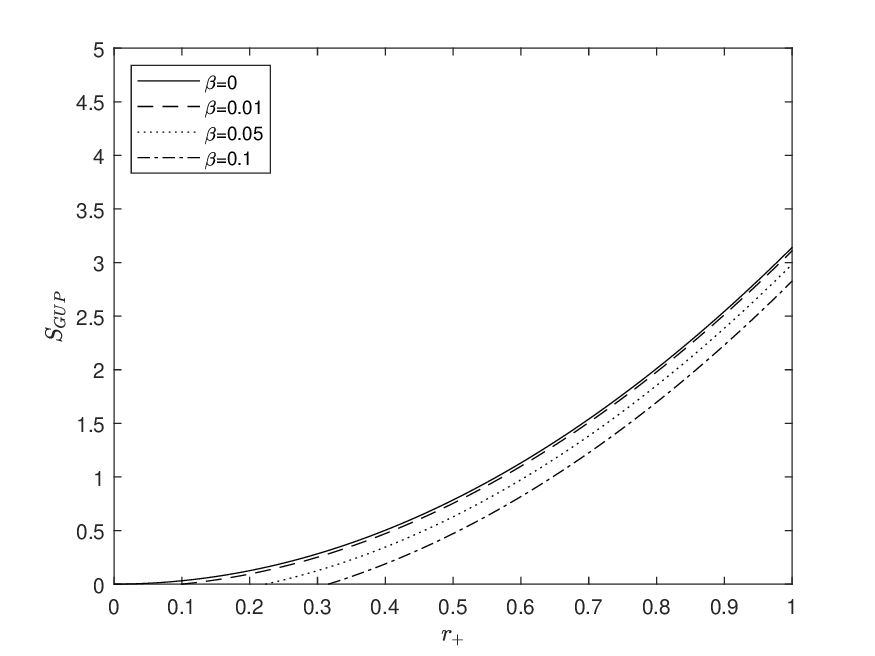}}
	\hfill
	\subfloat[]{\includegraphics[width=7cm]{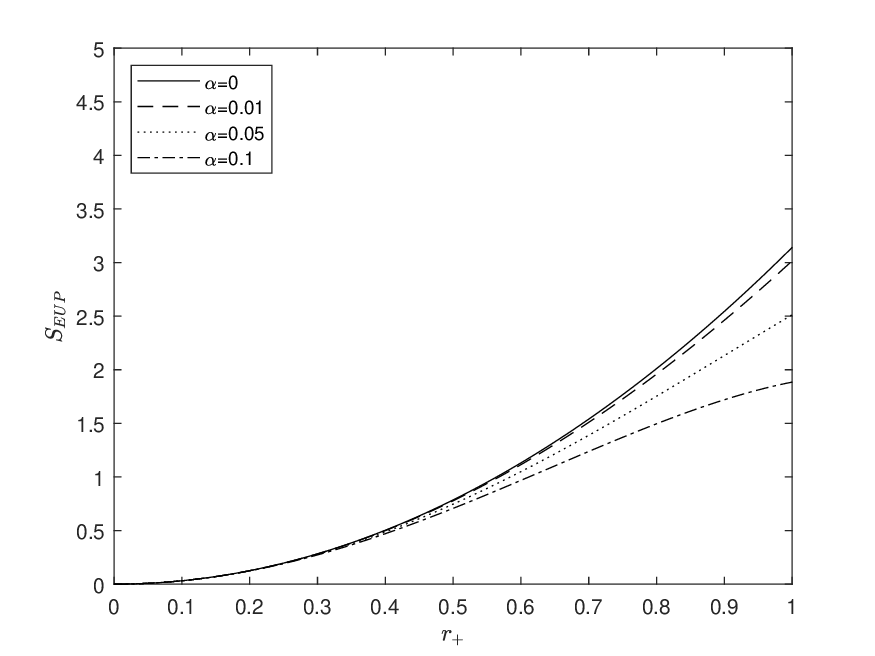}}
	\caption{\label{fig:5} Entropy versus event horizon for $l_p=L=1$. (a) EGUP correction of entropy. (b) GUP correction of entropy. (c) EUP correction of entropy.}
\end{figure}

According to the thermodynamics of black holes, we can find the entropy of black holes from the Hawking temperature. We can find the EGUP-modified entropy of the black hole:
\begin{align}
	S=&\int\frac{\mathrm{d}M}{T}\nonumber\\
	=&\frac{\pi\beta l_p^2}{4}\int\frac{\mathrm{d}r_+}{r_+\left[1-\sqrt{1-\beta l_p^2\left(\frac{1}{4r_+^2}+\frac{\alpha}{L^2}\right)}\right]}.\label{eq:twelve}
\end{align}
According to Eq. (\ref{eq:twelve}), black hole entropy is not modified by QM. It should be noted that in classical general relativity, the entropy of a black hole is proportional to its area. However, when considering quantum field theory effects, there will be corrections to this entropy. From the perspective of quantum field theory, near the event horizon of a black hole, fluctuations in quantum fields will have an impact on entropy. These quantum corrections mainly stem from the consideration of various microscopic degrees of freedom of quantum fields. These microscopic states are neglected in classical theories, but in the quantum case, they lead to corrections to the entropy of a black hole that are dominated by logarithmic terms. \cite{CHEN2022136994,anacleto2021noncommutative,anacleto2015quantum} But in order to get the analytical solution, we first do the Taylor expansion on Eq. (\ref{eq:twelve}), and then integrate, and get the following result:
\begin{equation}
	S_{EGUP}=S_{HUP}\left(1-\frac{4\alpha r_+^2}{L^2}-\frac{\beta l_p^2}{r_+^2}\right).
\end{equation}
Where $S_{HUP}=\pi r_+^2$. If $\alpha=0$ or $\beta=0$, the EUP and GUP modifications to the entropy of the black hole can be obtained:
\begin{align}
	S_{GUP}=&S_{HUP}\left(1-\frac{\beta l_p^2}{r_+^2}\right),\\
	S_{EUP}=&S_{HUP}\left(1-\frac{4\alpha r_+^2}{L^2}\right).
\end{align}
Similarly, in order to see more specifically the modification of the black hole entropy by EGUP, GUP, and EUP, we have plotted these cases in Fig. \ref{fig:5}. Whether it is GUP, EUP or EGUP, the entropy of the black hole is smaller after the correction than before the correction. It can be seen that GUP correction is more obvious in small scale, and EUP correction is more obvious in large scale. EGUP correction of black hole entropy is a combination of the two.

To this end, let's discuss the effects of QM and EGUP on pressure. It is well known that the pressure $P$ in AdS space-time has the following relationship with the radius of AdS space-time $l$:
\begin{equation}
	P=\frac{3}{8\pi l^2}.\label{eq:thirteen}
\end{equation}
We can solve for $l^2$ from Eq.~(\ref{eq:eight}). After substituting Eq.~(\ref{eq:thirteen}), the pressure of EGUP and QM on AdS space-time can be corrected:
\begin{align}
	P_{EGUP}=&\frac{T}{\frac{16r_+^3}{\beta l_p^2}\left[1-\sqrt{1-\beta l_p^2\left(\frac{1}{4r_+^2}+\frac{\alpha}{L^2}\right)}\right]}\nonumber\\
	&+\frac{Q^2}{8\pi r_+^2}-\frac{1}{8\pi r_+^2}-\frac{3\omega_q\chi}{8\pi r_+^{3\omega_q+3}}.
\end{align}
For $\alpha=0$, the GUP correction to the pressure is obtained:
\begin{equation}
	P_{GUP}=\frac{T}{\frac{16r_+^3}{\beta l_p^2}\left[1-\sqrt{1-\frac{\beta l_p^2}{4r_+^2}}\right]}+\frac{Q^2}{8\pi r_+^2}-\frac{1}{8\pi r_+^2}-\frac{3\omega_q\chi}{8\pi r_+^{3\omega_q+3}}.
\end{equation}
For $\beta=0$, the EUP correction to the pressure is obtained:
\begin{equation}
	P_{EUP}=\frac{T}{8r_+^3\left(\frac{1}{4r_+^2}+\frac{\alpha}{L^2}\right)}+\frac{Q^2}{8\pi r_+^2}-\frac{1}{8\pi r_+^2}-\frac{3\omega_q\chi}{8\pi r_+^{3\omega_q+3}}.
\end{equation}
For $\alpha=0$ and $\beta=0$, the HUP correction to the pressure is obtained:
\begin{equation}
	P_{HUP}=\frac{T}{2r_+}+\frac{Q^2}{8\pi r_+^2}-\frac{1}{8\pi r_+^2}-\frac{3\omega_q\chi}{8\pi r_+^{3\omega_q+3}}.
\end{equation}
\begin{figure}[!h]
	\subfloat[]{\includegraphics[width=7cm]{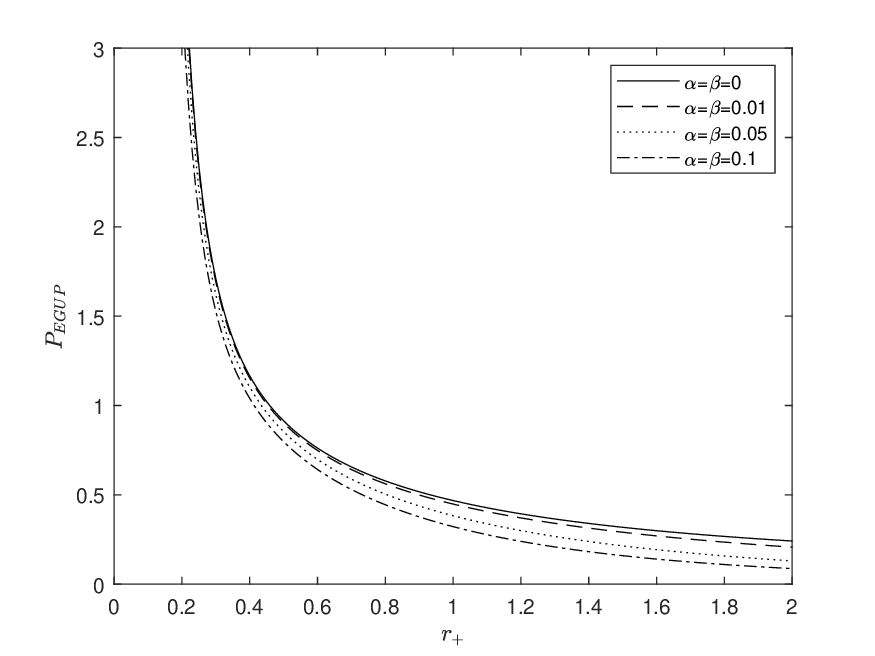}}
	\hfill
	\subfloat[]{\includegraphics[width=7cm]{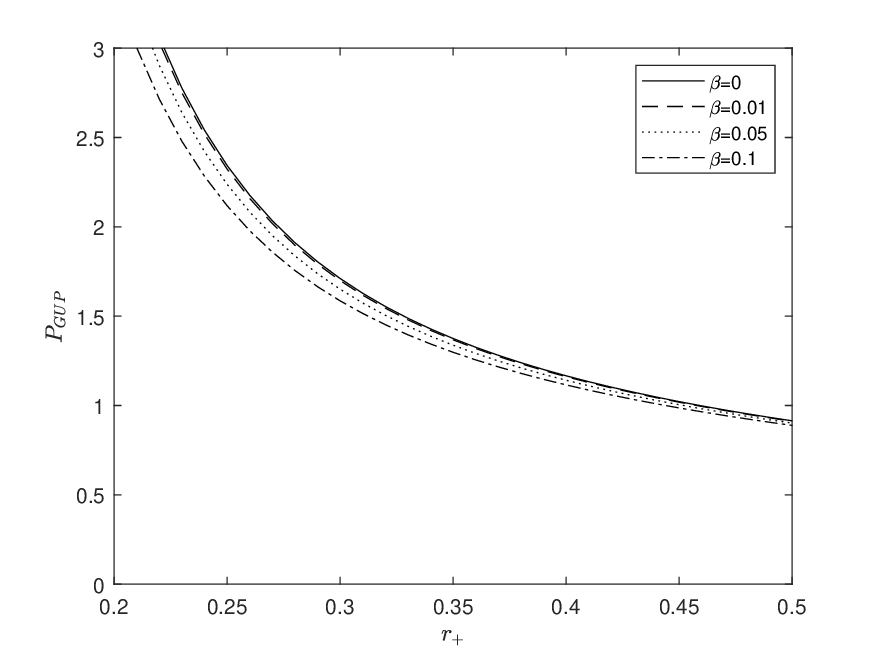}}
	\hfill
	\subfloat[]{\includegraphics[width=7cm]{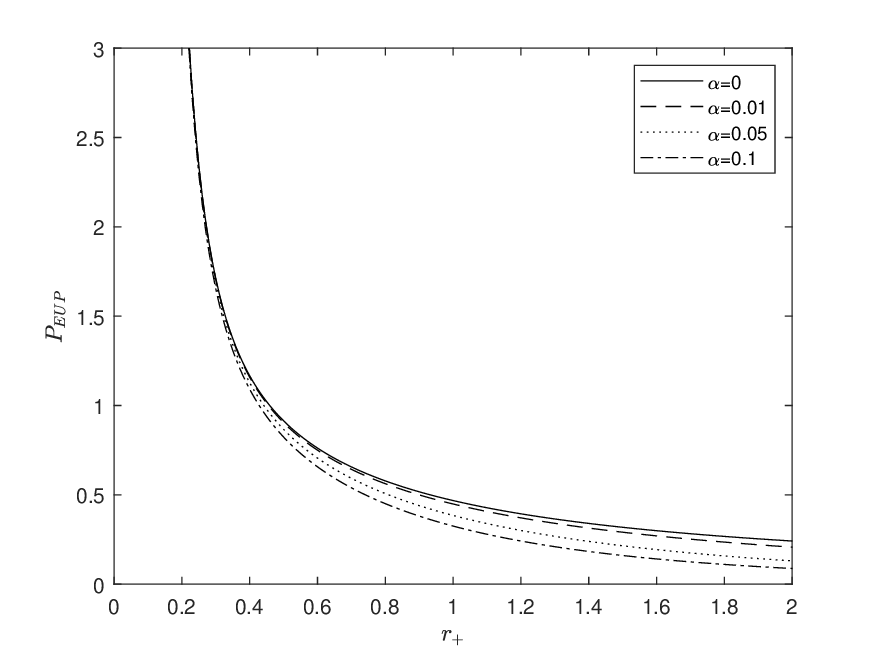}}
	\caption{\label{fig:6}  Pressure versus event horizon for $l_p=L=1$, $T=1$, $Q=0.3$, $\chi=0.1$, $\omega_q=-1/3$. (a) EGUP correction of  pressure. (b) GUP correction of pressure. (c) EUP correction of pressure.}
\end{figure}
\begin{figure}[!h]
	\subfloat[]{\includegraphics[width=7cm]{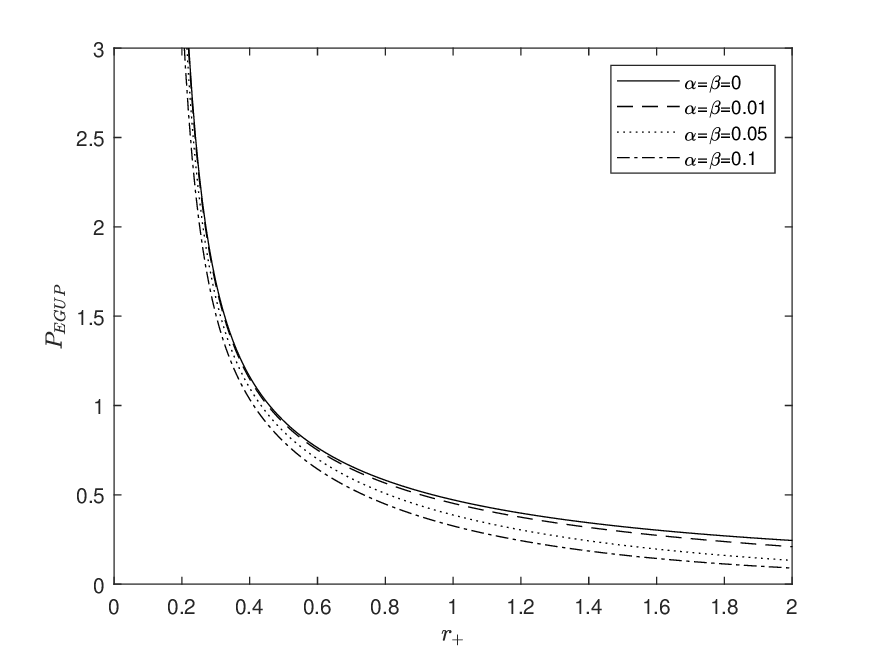}}
	\hfill
	\subfloat[]{\includegraphics[width=7cm]{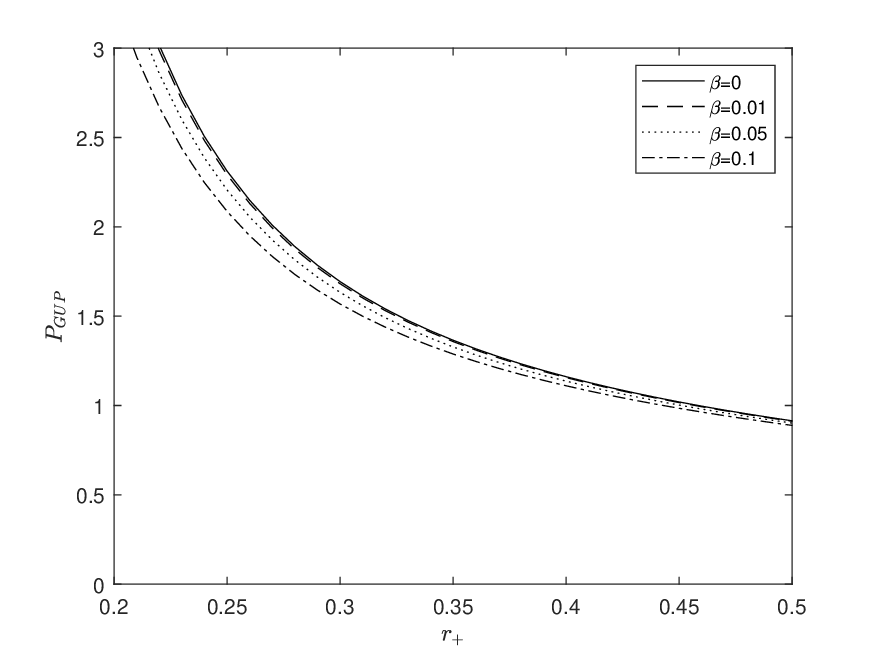}}
	\hfill
	\subfloat[]{\includegraphics[width=7cm]{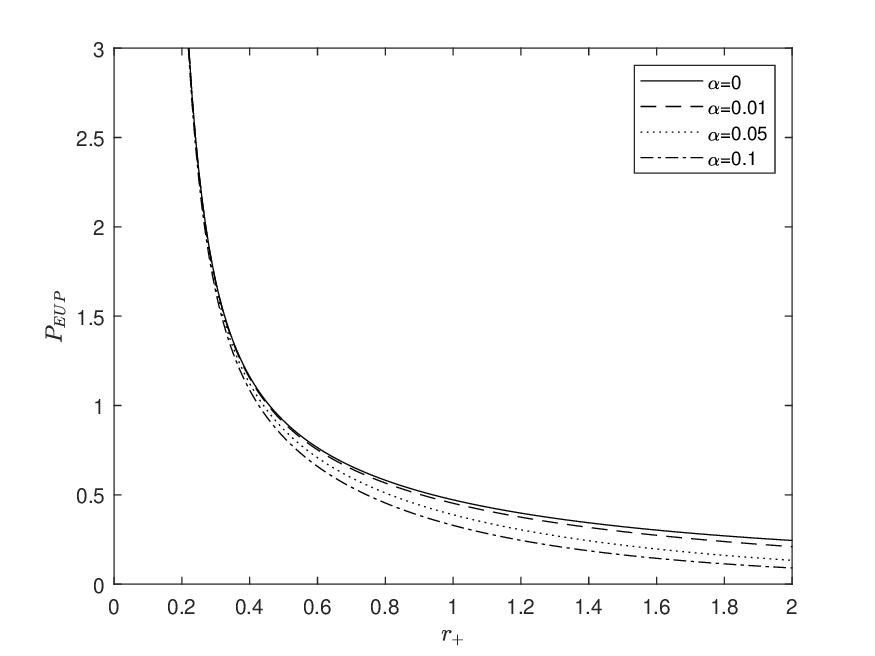}}
	\caption{\label{fig:7}  Pressure versus event horizon for $l_p=L=1$, $T=1$, $Q=0.3$, $\chi=0.1$, $\omega_q=-2/3$. (a) EGUP correction of pressure. (b) GUP correction of pressure. (c) EUP correction of pressure.}
\end{figure}

To more clearly show the relationship between the pressure and the location of the black hole's event horizon, we plot the $\omega_q=-1/3$ case and the $\omega_q=-2/3$ case in Fig. \ref{fig:6} and Fig. \ref{fig:7}, respectively. Similarly, we found that for pressure, whether $\omega_q=-1/3$ or $\omega_q=-2/3$ is chosen, there is only a slight correction to the pressure without altering the overall trend. Whether it is GUP, EUP, or EGUP correction, it will lead to a decrease in pressure. Similar to what was mentioned earlier, the correction of GUP to pressure is mainly concentrated on small scales, while it has almost no effect on large scales. The correction of EUP to pressure is mainly concentrated on large scales, while it has almost no effect on small scales. The correction of EGUP to pressure is a combination of the two, affecting both large and small scales.
\section{Conclusion}
In this paper, we discuss the modification of the thermodynamics of RN-AdS black holes by EGUP and QM. First, we discuss the modification of Hawking temperature by EGUP and QM. And in order to ensure that the temperature is real and positive, we find that the event horizon of the black hole appears a minimum radius under the modification of GUP and EGUP. In addition, the QM term also leads to the appearance of the minimum radius of the black hole. then we studied the modifications of EGUP and QM to other thermodynamic quantities such as pressure, entropy, heat capacity, and calculated them in the EUP and GUP limits. We find that the main difference between EUP and GUP on the thermodynamic quantity correction of black hole is that the main effect of EUP correction is reflected in the large scale region, while the effect of GUP correction is mainly reflected in the small scale region. To this end, we find that although RN black holes and RN-AdS black holes seem to be little different from each other in terms of linear elements, their thermodynamic properties have essentially changed. First, for the QM term of an RN-AdS black hole, there is no qualitative difference between $\omega_q=-1/3$ and $\omega_q=-2/3$, only a small numerical difference. Second, the phase transition point of RN-AdS black holes shift to nearly zero. The heat capacity of black holes is negative only in a small range, and is positive almost everywhere, resulting in the stable existence of low-mass black holes. Both of these are very different from RN black holes. The reason for almost shift to zero of phase transition points of RN-AdS black holes remains to be further studied.

\section*{Acknowledgements}
The authors are grateful to anonymous reviewers for their very crucial and detailed comments. The authors would like to thank Prof. Jian Jing and his student LiuBiao Ma from the College of Mathematics and Physics , Beijing University of Chemical Technology for their valuable comments and suggestions during the completion of this manuscript.
  \bibliographystyle{elsarticle-num} 
  \bibliography{EGUPQM}

\begin{thebibliography}{10}
\expandafter\ifx\csname url\endcsname\relax
  \def\url#1{\texttt{#1}}\fi
\expandafter\ifx\csname urlprefix\endcsname\relax\def\urlprefix{URL }\fi
\expandafter\ifx\csname href\endcsname\relax
  \def\href#1#2{#2} \def\path#1{#1}\fi

\bibitem{PhysRevD.7.2333}
J.~D. Bekenstein, \href{https://link.aps.org/doi/10.1103/PhysRevD.7.2333}{Black
  holes and entropy}, Phys. Rev. D 7~(8) (1973) 2333--2346.
\newblock \href {https://doi.org/10.1103/PhysRevD.7.2333}
  {\path{doi:10.1103/PhysRevD.7.2333}}.
\newline\urlprefix\url{https://link.aps.org/doi/10.1103/PhysRevD.7.2333}

\bibitem{hawking1975particle}
S.~Hawking, \href{https://doi.org/10.1007/BF02345020}{Particle creation by
  black holes}, Commun.Math. Phys. 43~(3) (1975) 199--220.
\newblock \href {https://doi.org/10.1007/BF02345020}
  {\path{doi:10.1007/BF02345020}}.
\newline\urlprefix\url{https://doi.org/10.1007/BF02345020}

\bibitem{doi:10.1142/S0217751X95000085}
L.~J. GARAY, \href{https://doi.org/10.1142/S0217751X95000085}{Quantum gravity
  and minimum length}, International Journal of Modern Physics A 10~(02) (1995)
  145--165.
\newblock \href {https://doi.org/10.1142/S0217751X95000085}
  {\path{doi:10.1142/S0217751X95000085}}.
\newline\urlprefix\url{https://doi.org/10.1142/S0217751X95000085}

\bibitem{KONISHI1990276}
K.~Konishi, G.~Paffuti, P.~Provero,
  \href{https://doi.org/10.1016/0370-2693(90)91927-4}{Minimum physical length
  and the generalized uncertainty principle in string theory}, Physics Letters
  B 234~(3) (1990) 276--284.
\newblock \href {https://doi.org/10.1016/0370-2693(90)91927-4}
  {\path{doi:10.1016/0370-2693(90)91927-4}}.
\newline\urlprefix\url{https://doi.org/10.1016/0370-2693(90)91927-4}

\bibitem{capozziello2000generalized}
S.~Capozziello, G.~Lambiase, G.~Scarpetta,
  \href{https://doi.org/10.1023/A:1003634814685}{Generalized uncertainty
  principle from quantumgeometry}, International Journal of Theoretical Physics
  39~(1) (2000) 15--22.
\newblock \href {https://doi.org/10.1023/A:1003634814685}
  {\path{doi:10.1023/A:1003634814685}}.
\newline\urlprefix\url{https://doi.org/10.1023/A:1003634814685}

\bibitem{hooft2009dimensionalreductionquantumgravity}
G.~'t~Hooft, \href{https://arxiv.org/abs/gr-qc/9310026}{Dimensional reduction
  in quantum gravity} (2009).
\newblock \href {http://arxiv.org/abs/gr-qc/9310026}
  {\path{arXiv:gr-qc/9310026}}.
\newline\urlprefix\url{https://arxiv.org/abs/gr-qc/9310026}

\bibitem{susskind1995world}
L.~Susskind, \href{https://doi.org/10.1063/1.531249}{The world as a hologram},
  J. Math. Phys. 36~(11) (1995) 6377--6396.
\newblock \href {https://doi.org/10.1063/1.531249}
  {\path{doi:10.1063/1.531249}}.
\newline\urlprefix\url{https://doi.org/10.1063/1.531249}

\bibitem{maldacena1999large}
J.~Maldacena, \href{https://doi.org/10.1023/A:1026654312961}{The large-n limit
  of superconformal field theories and supergravity}, International journal of
  theoretical physics 38~(4) (1999) 1113--1133.
\newblock \href {https://doi.org/10.1023/A:1026654312961}
  {\path{doi:10.1023/A:1026654312961}}.
\newline\urlprefix\url{https://doi.org/10.1023/A:1026654312961}

\bibitem{PhysRevLett.126.101601}
A.~Al~Balushi, R.~A. Hennigar, H.~K. Kunduri, R.~B. Mann,
  \href{https://link.aps.org/doi/10.1103/PhysRevLett.126.101601}{Holographic
  complexity and thermodynamic volume}, Phys. Rev. Lett. 126~(10) (2021)
  101601.
\newblock \href {https://doi.org/10.1103/PhysRevLett.126.101601}
  {\path{doi:10.1103/PhysRevLett.126.101601}}.
\newline\urlprefix\url{https://link.aps.org/doi/10.1103/PhysRevLett.126.101601}

\bibitem{gwak2017thermodynamics}
B.~Gwak, \href{https://doi.org/10.1007/JHEP11(2017)129}{Thermodynamics with
  pressure and volume under charged particle absorption}, Journal of High
  Energy Physics 2017~(11) (2017) 1--16.
\newblock \href {https://doi.org/10.1007/JHEP11(2017)129}
  {\path{doi:10.1007/JHEP11(2017)129}}.
\newline\urlprefix\url{https://doi.org/10.1007/JHEP11(2017)129}

\bibitem{RevModPhys.95.035003}
D.~Harlow, B.~Heidenreich, M.~Reece, T.~Rudelius,
  \href{https://link.aps.org/doi/10.1103/RevModPhys.95.035003}{Weak gravity
  conjecture}, Rev. Mod. Phys. 95~(3) (2023) 035003.
\newblock \href {https://doi.org/10.1103/RevModPhys.95.035003}
  {\path{doi:10.1103/RevModPhys.95.035003}}.
\newline\urlprefix\url{https://link.aps.org/doi/10.1103/RevModPhys.95.035003}

\bibitem{doi:10.1142/S021827180600942X}
E.~J. COPELAND, M.~SAMI, S.~TSUJIKAWA,
  \href{https://doi.org/10.1142/S021827180600942X}{Dynamics of dark energy},
  International Journal of Modern Physics D 15~(11) (2006) 1753--1935.
\newblock \href {https://doi.org/10.1142/S021827180600942X}
  {\path{doi:10.1142/S021827180600942X}}.
\newline\urlprefix\url{https://doi.org/10.1142/S021827180600942X}

\bibitem{bahcall1999cosmic}
N.~A. Bahcall, J.~P. Ostriker, S.~Perlmutter, P.~J. Steinhardt,
  \href{https://www.science.org/doi/10.1126/science.284.5419.1481}{The cosmic
  triangle: Revealing the state of the universe}, science 284~(5419) (1999)
  1481--1488.
\newblock \href {https://doi.org/10.1126/science.284.5419.1481}
  {\path{doi:10.1126/science.284.5419.1481}}.
\newline\urlprefix\url{https://www.science.org/doi/10.1126/science.284.5419.1481}

\bibitem{PhysRevD.59.123504}
P.~J. Steinhardt, L.~Wang, I.~Zlatev,
  \href{https://link.aps.org/doi/10.1103/PhysRevD.59.123504}{Cosmological
  tracking solutions}, Phys. Rev. D 59~(12) (1999) 123504.
\newblock \href {https://doi.org/10.1103/PhysRevD.59.123504}
  {\path{doi:10.1103/PhysRevD.59.123504}}.
\newline\urlprefix\url{https://link.aps.org/doi/10.1103/PhysRevD.59.123504}

\bibitem{kiselev2003quintessence}
V.~Kiselev,
  \href{https://iopscience.iop.org/article/10.1088/0264-9381/20/6/310}{Quintessence
  and black holes}, Classical and Quantum Gravity 20~(6) (2003) 1187.
\newblock \href {https://doi.org/10.1088/0264-9381/20/6/310}
  {\path{doi:10.1088/0264-9381/20/6/310}}.
\newline\urlprefix\url{https://iopscience.iop.org/article/10.1088/0264-9381/20/6/310}

\bibitem{yan2021hawking}
D.-W. Yan, Z.-R. Huang, N.~Li,
  \href{https://iopscience.iop.org/article/10.1088/1674-1137/abc0cf}{Hawking-page
  phase transitions of charged ads black holes surrounded by quintessence},
  Chinese Physics C 45~(1) (2021) 015104.
\newblock \href {https://doi.org/10.1088/1674-1137/abc0cf}
  {\path{doi:10.1088/1674-1137/abc0cf}}.
\newline\urlprefix\url{https://iopscience.iop.org/article/10.1088/1674-1137/abc0cf}

\bibitem{huang2021phase}
Y.~Huang, H.~Jing, J.~Tao, F.~Yao,
  \href{https://iopscience.iop.org/article/10.1088/1674-1137/abf6c4}{Phase
  structures and transitions of quintessence surrounding rn black holes in a
  grand canonical ensemble}, Chinese Physics C 45~(7) (2021) 075101.
\newblock \href {https://doi.org/10.1088/1674-1137/abf6c4}
  {\path{doi:10.1088/1674-1137/abf6c4}}.
\newline\urlprefix\url{https://iopscience.iop.org/article/10.1088/1674-1137/abf6c4}

\bibitem{lutfuouglu2021thermodynamics}
B.~L{\"u}tf{\"u}o{\u{g}}lu, B.~Hamil, L.~Dahbi,
  \href{https://doi.org/10.1140/epjp/s13360-021-01975-y}{Thermodynamics of
  schwarzschild black hole surrounded by quintessence with generalized
  uncertainty principle}, The European Physical Journal Plus 136~(9) (2021)
  976.
\newblock \href {https://doi.org/10.1140/epjp/s13360-021-01975-y}
  {\path{doi:10.1140/epjp/s13360-021-01975-y}}.
\newline\urlprefix\url{https://doi.org/10.1140/epjp/s13360-021-01975-y}

\bibitem{CHEN2022136994}
H.~Chen, B.~C. Lütfüoğlu, H.~Hassanabadi, Z.-W. Long,
  \href{https://doi.org/10.1016/j.physletb.2022.136994}{Thermodynamics of the
  reissner-nordström black hole with quintessence matter on the egup
  framework}, Physics Letters B 827 (2022) 136994.
\newblock \href {https://doi.org/10.1016/j.physletb.2022.136994}
  {\path{doi:10.1016/j.physletb.2022.136994}}.
\newline\urlprefix\url{https://doi.org/10.1016/j.physletb.2022.136994}

\bibitem{PARK2008698}
M.-I. Park, \href{https://doi.org/10.1016/j.physletb.2007.11.090}{The
  generalized uncertainty principle in (a)ds space and the modification of
  hawking temperature from the minimal length}, Physics Letters B 659~(3)
  (2008) 698--702.
\newblock \href {https://doi.org/10.1016/j.physletb.2007.11.090}
  {\path{doi:10.1016/j.physletb.2007.11.090}}.
\newline\urlprefix\url{https://doi.org/10.1016/j.physletb.2007.11.090}

\bibitem{doi:10.1142/S0218271814300250}
A.~Tawfik, A.~Diab,
  \href{https://doi.org/10.1142/S0218271814300250}{Generalized uncertainty
  principle: Approaches and applications}, International Journal of Modern
  Physics D 23~(12) (2014) 1430025.
\newblock \href {https://doi.org/10.1142/S0218271814300250}
  {\path{doi:10.1142/S0218271814300250}}.
\newline\urlprefix\url{https://doi.org/10.1142/S0218271814300250}

\bibitem{FENG201781}
Z.-W. Feng, S.-Z. Yang, H.-L. Li, X.-T. Zu,
  \href{https://doi.org/10.1016/j.physletb.2017.02.043}{Constraining the
  generalized uncertainty principle with the gravitational wave event
  gw150914}, Physics Letters B 768 (2017) 81--85.
\newblock \href {https://doi.org/10.1016/j.physletb.2017.02.043}
  {\path{doi:10.1016/j.physletb.2017.02.043}}.
\newline\urlprefix\url{https://doi.org/10.1016/j.physletb.2017.02.043}

\bibitem{doi:10.1142/S0218271817500626}
S.~Zhou, G.-R. Chen, \href{https://doi.org/10.1142/S0218271817500626}{Corrected
  black hole thermodynamics in damour–ruffini’s method with generalized
  uncertainty principle}, International Journal of Modern Physics D 26~(07)
  (2017) 1750062.
\newblock \href {https://doi.org/10.1142/S0218271817500626}
  {\path{doi:10.1142/S0218271817500626}}.
\newline\urlprefix\url{https://doi.org/10.1142/S0218271817500626}

\bibitem{PEDRAM2012638}
P.~Pedram, \href{https://doi.org/10.1016/j.physletb.2012.10.059}{A higher order
  gup with minimal length uncertainty and maximal momentum ii: Applications},
  Physics Letters B 718~(2) (2012) 638--645.
\newblock \href {https://doi.org/10.1016/j.physletb.2012.10.059}
  {\path{doi:10.1016/j.physletb.2012.10.059}}.
\newline\urlprefix\url{https://doi.org/10.1016/j.physletb.2012.10.059}

\bibitem{ali2011minimal}
A.~F. Ali,
  \href{https://iopscience.iop.org/article/10.1088/0264-9381/28/6/065013}{Minimal
  length in quantum gravity, equivalence principle and holographic entropy
  bound}, Classical and Quantum Gravity 28~(6) (2011) 065013.
\newblock \href {https://doi.org/10.1088/0264-9381/28/6/065013}
  {\path{doi:10.1088/0264-9381/28/6/065013}}.
\newline\urlprefix\url{https://iopscience.iop.org/article/10.1088/0264-9381/28/6/065013}

\bibitem{nozari2010minimal}
K.~Nozari, P.~Pedram,
  \href{https://iopscience.iop.org/article/10.1209/0295-5075/92/50013}{Minimal
  length and bouncing-particle spectrum}, Europhysics Letters 92~(5) (2010)
  50013.
\newblock \href {https://doi.org/10.1209/0295-5075/92/50013}
  {\path{doi:10.1209/0295-5075/92/50013}}.
\newline\urlprefix\url{https://iopscience.iop.org/article/10.1209/0295-5075/92/50013}

\bibitem{PhysRevD.74.104001}
W.~Kim, Y.-W. Kim, Y.-J. Park,
  \href{https://link.aps.org/doi/10.1103/PhysRevD.74.104001}{Entropy of the
  randall-sundrum brane world with the generalized uncertainty principle},
  Phys. Rev. D 74~(10) (2006) 104001.
\newblock \href {https://doi.org/10.1103/PhysRevD.74.104001}
  {\path{doi:10.1103/PhysRevD.74.104001}}.
\newline\urlprefix\url{https://link.aps.org/doi/10.1103/PhysRevD.74.104001}

\bibitem{setare2006generalized}
M.~Setare,
  \href{https://www.worldscientific.com/doi/abs/10.1142/S0217751X06025304}{The
  generalized uncertainty principle and corrections to the cardy--verlinde
  formula in sads 5 black holes}, International Journal of Modern Physics A
  21~(06) (2006) 1325--1332.
\newblock \href {https://doi.org/10.1142/S0217751X06025304}
  {\path{doi:10.1142/S0217751X06025304}}.
\newline\urlprefix\url{https://www.worldscientific.com/doi/abs/10.1142/S0217751X06025304}

\bibitem{maghsoodi2020effect}
E.~Maghsoodi, H.~Hassanabadi, W.~S. Chung,
  \href{https://iopscience.iop.org/article/10.1209/0295-5075/129/59001}{Effect
  of the new extended uncertainty principle on black hole thermodynamics},
  Europhysics Letters 129~(5) (2020) 59001.
\newblock \href {https://doi.org/10.1209/0295-5075/129/59001}
  {\path{doi:10.1209/0295-5075/129/59001}}.
\newline\urlprefix\url{https://iopscience.iop.org/article/10.1209/0295-5075/129/59001}

\bibitem{anacleto2021noncommutative}
M.~Anacleto, F.~Brito, S.~Cruz, E.~Passos,
  \href{https://doi.org/10.1142/S0217751X21500287}{Noncommutative correction to
  the entropy of schwarzschild black hole with gup}, International Journal of
  Modern Physics A 36~(03) (2021) 2150028.
\newblock \href {https://doi.org/10.1142/S0217751X21500287}
  {\path{doi:10.1142/S0217751X21500287}}.
\newline\urlprefix\url{https://doi.org/10.1142/S0217751X21500287}

\bibitem{doi:10.1142/S0217732310033426}
S.~MIGNEMI, \href{https://doi.org/10.1142/S0217732310033426}{Extended
  uncertainty principle and the geometry of (anti)-de sitter space}, Modern
  Physics Letters A 25~(20) (2010) 1697--1703.
\newblock \href {https://doi.org/10.1142/S0217732310033426}
  {\path{doi:10.1142/S0217732310033426}}.
\newline\urlprefix\url{https://doi.org/10.1142/S0217732310033426}

\bibitem{CHUNG2019451}
W.~S. Chung, H.~Hassanabadi,
  \href{https://doi.org/10.1016/j.physletb.2019.04.063}{Black hole temperature
  and unruh effect from the extended uncertainty principle}, Physics Letters B
  793 (2019) 451--456.
\newblock \href {https://doi.org/10.1016/j.physletb.2019.04.063}
  {\path{doi:10.1016/j.physletb.2019.04.063}}.
\newline\urlprefix\url{https://doi.org/10.1016/j.physletb.2019.04.063}

\bibitem{hamil2021effect}
B.~Hamil, B.~L{\"u}tf{\"u}o{\u{g}}lu,
  \href{https://iopscience.iop.org/article/10.1209/0295-5075/133/30003}{Effect
  of the modified heisenberg algebra on the black hole thermodynamics},
  Europhysics Letters 133~(3) (2021) 30003.
\newblock \href {https://doi.org/10.1209/0295-5075/133/30003}
  {\path{doi:10.1209/0295-5075/133/30003}}.
\newline\urlprefix\url{https://iopscience.iop.org/article/10.1209/0295-5075/133/30003}

\bibitem{hamil2021effect1}
B.~Hamil, B.~L{\"u}tf{\"u}o{\u{g}}lu,
  \href{https://iopscience.iop.org/article/10.1209/0295-5075/134/50007}{The
  effect of higher-order extended uncertainty principle on the black hole
  thermodynamics}, Europhysics Letters 134~(5) (2021) 50007.
\newblock \href {https://doi.org/10.1209/0295-5075/134/50007}
  {\path{doi:10.1209/0295-5075/134/50007}}.
\newline\urlprefix\url{https://iopscience.iop.org/article/10.1209/0295-5075/134/50007}

\bibitem{hamil2021black}
B.~Hamil, B.~C. L{\"u}tf{\"u}o{\u{g}}lu,
  \href{https://iopscience.iop.org/article/10.1209/0295-5075/135/59001}{Black
  hole thermodynamics in the presence of a maximal length and minimum
  measurable in momentum}, Europhysics Letters 135~(5) (2021) 59001.
\newblock \href {https://doi.org/10.1209/0295-5075/135/59001}
  {\path{doi:10.1209/0295-5075/135/59001}}.
\newline\urlprefix\url{https://iopscience.iop.org/article/10.1209/0295-5075/135/59001}

\bibitem{bolen2005anti}
B.~Bolen, M.~Cavaglia, \href{https://doi.org/10.1007/s10714-005-0108-x}{(anti-)
  de sitter black hole thermodynamics and the generalized uncertainty
  principle}, General Relativity and Gravitation 37 (2005) 1255--1262.
\newblock \href {https://doi.org/10.1007/s10714-005-0108-x}
  {\path{doi:10.1007/s10714-005-0108-x}}.
\newline\urlprefix\url{https://doi.org/10.1007/s10714-005-0108-x}

\bibitem{doi:10.1142/S0217732321502023}
M.~Sadeghi, \href{https://doi.org/10.1142/S0217732321502023}{Ads black brane
  solution surrounded by quintessence in massive gravity and kss bound}, Modern
  Physics Letters A 36~(28) (2021) 2150202.
\newblock \href {https://doi.org/10.1142/S0217732321502023}
  {\path{doi:10.1142/S0217732321502023}}.
\newline\urlprefix\url{https://doi.org/10.1142/S0217732321502023}

\bibitem{ghosh2018lovelock}
S.~G. Ghosh, S.~D. Maharaj, D.~Baboolal, T.-H. Lee,
  \href{https://doi.org/10.1140/epjc/s10052-018-5570-1}{Lovelock black holes
  surrounded by quintessence}, The European Physical Journal C 78 (2018) 1--8.
\newblock \href {https://doi.org/10.1140/epjc/s10052-018-5570-1}
  {\path{doi:10.1140/epjc/s10052-018-5570-1}}.
\newline\urlprefix\url{https://doi.org/10.1140/epjc/s10052-018-5570-1}

\bibitem{PhysRevD.91.123002}
C.~G. B\"ohmer, N.~Tamanini, M.~Wright,
  \href{https://link.aps.org/doi/10.1103/PhysRevD.91.123002}{Interacting
  quintessence from a variational approach. i. algebraic couplings}, Phys. Rev.
  D 91~(12) (2015) 123002.
\newblock \href {https://doi.org/10.1103/PhysRevD.91.123002}
  {\path{doi:10.1103/PhysRevD.91.123002}}.
\newline\urlprefix\url{https://link.aps.org/doi/10.1103/PhysRevD.91.123002}

\bibitem{xiang2009heuristic}
L.~Xiang, X.~Wen,
  \href{https://iopscience.iop.org/article/10.1088/1126-6708/2009/10/046}{A
  heuristic analysis of black hole thermodynamics with generalized uncertainty
  principle}, Journal of High Energy Physics 2009~(10) (2009) 046.
\newblock \href {https://doi.org/10.1088/1126-6708/2009/10/046}
  {\path{doi:10.1088/1126-6708/2009/10/046}}.
\newline\urlprefix\url{https://iopscience.iop.org/article/10.1088/1126-6708/2009/10/046}

\bibitem{PhysRevD.70.124021}
A.~J.~M. Medved, E.~C. Vagenas,
  \href{https://link.aps.org/doi/10.1103/PhysRevD.70.124021}{When conceptual
  worlds collide: The generalized uncertainty principle and the
  bekenstein-hawking entropy}, Phys. Rev. D 70~(12) (2004) 124021.
\newblock \href {https://doi.org/10.1103/PhysRevD.70.124021}
  {\path{doi:10.1103/PhysRevD.70.124021}}.
\newline\urlprefix\url{https://link.aps.org/doi/10.1103/PhysRevD.70.124021}

\bibitem{anacleto2015quantum}
M.~Anacleto, F.~Brito, E.~Passos,
  \href{https://doi.org/10.1016/j.physletb.2015.07.072}{Quantum-corrected
  self-dual black hole entropy in tunneling formalism with gup}, Physics
  Letters B 749 (2015) 181--186.
\newblock \href {https://doi.org/10.1016/j.physletb.2015.07.072}
  {\path{doi:10.1016/j.physletb.2015.07.072}}.
\newline\urlprefix\url{https://doi.org/10.1016/j.physletb.2015.07.072}

\end{thebibliography}

\end{document}